\documentclass[journal]{IEEEtran}
\usepackage{amsmath,amsfonts}
\usepackage{algorithmic}
\usepackage{algorithm}
\usepackage{array}
\usepackage[caption=false,font=normalsize,labelfont=sf,textfont=sf]{subfig}
\usepackage{textcomp}
\usepackage{stfloats}
\usepackage{url}
\usepackage{verbatim}
\usepackage{graphicx}
\usepackage{cite}
\usepackage{mathtools}
\usepackage{hyperref}
\usepackage{amsmath}
\begin{document}

\title{A Multi-Modal Latent-Features based Service Recommendation System for the Social Internet of Things}

\author{
Amar Khelloufi, Huansheng Ning ~\IEEEmembership{Senior Member,~IEEE}, Abdenacer Naouri, Abdelkarim Ben Sada, Attia Qammar, Abdelkader Khalil, Sahraoui Dhelim and Lingfeng Mao.

\thanks{A. Khelloufi, H. Ning, A. Naouri, A. Ben Sada, A. Qammar and Lingfeng Mao are with the School of Computer and Communication Engineering, University of Science and Technology Beijing, Beijing 10083, China, email:amar.khelloufi@hotmail.com, ninghuansheng@ustb.edu.cn, nacer.naouri@gmail.com, ab\_ben@live.com, q.attia@yahoo.com}

\thanks{S. Dhelim is with the School of Computer Science, University College Dublin, Ireland, Corresponding author: Sahraoui Dhelim (email: sahraoui.dhelim@ucd.ie)}
\thanks{A. Khalil is with Ziane Achour University of Djelfa, Algeria} 
\thanks{Manuscript received Month Day, Year; revised Month Day, Year.}
}

\markboth{Journal of \LaTeX\ Class Files,~Vol.~14, No.~8, August~2021}%
{Shell \MakeLowercase{\textit{et al.}}: A Sample Article Using IEEEtran.cls for IEEE Journals}


\maketitle

\begin{abstract}
The Social Internet of Things (SIoT), is revolutionizing how we interact with our everyday lives. By adding the social dimension to connecting devices, the SIoT has the potential to drastically change the way we interact with smart devices. This connected infrastructure allows for unprecedented levels of convenience, automation, and access to information, allowing us to do more with less effort. However, this revolutionary new technology also brings an eager need for service recommendation systems. As the SIoT grows in scope and complexity, it becomes increasingly important for businesses and individuals, and SIoT objects alike to have reliable sources for products, services, and information that are tailored to their specific needs. Few works have been proposed to provide service recommendations for SIoT environments. However, these efforts have been confined to only focusing on modeling user-item interactions using contextual information, devices’ SIoT relationships, and correlation social groups but these schemes do not account for latent semantic item-item structures underlying the sparse multi-modal contents in SIoT environment. In this paper, we propose a latent-based SIoT recommendation system that learns item-item structures and aggregates multiple modalities to obtain latent item graphs which are then used in graph convolutions to inject high-order affinities into item representations. Experiments showed that the proposed recommendation system outperformed state-of-the-art SIoT recommendation methods and validated its efficacy at mining latent relationships from multi-modal features. 
\end{abstract}


\section{Introduction}
\label{sec:introduction}
\IEEEPARstart{T}{he} Internet of Things (IoT) is the transformational paradigm that connects people, devices, and things, redefining how the physical world interacts with the digital one. It provides a network for 'things' such as sensors and actuators to be connected and monitored, creating a massive amount of data that needs to be both produced and consumed in order to create an integrated environment \cite{ashton2009internet}. As technology advances, these IoT devices are becoming increasingly intelligent as well as capable of offering special services. According to research estimates, by 2024 there will be approximately 20 billion devices connected to the IoT network \cite{al2020internet}.  The IoT is revolutionizing how we interact with the world around us, providing access to data and analytics that would have been completely inaccessible before. This technology has broad implications for both businesses and individuals, allowing them to see gains in efficiency, safety measures, cost savings, and more. Companies that recognize the value of this technology can stay one step ahead of their competitors by investing in it early on. The Social Internet of Things (SIoT), on the other hand, is a major breakthrough in how we interact with everyday objects. By allowing devices to form social relationships, SIoT will revolutionize our lives and the way we do things. For example, if your fridge and washing machine are connected to SIoT, they could communicate and coordinate together to ensure that you always have clean clothes and food in your home. The possibilities are endless, and this technology will undoubtedly revolutionize the way we live. With SIoT, tasks such as ordering groceries or scheduling maintenance on appliances can become much easier, saving us time and money. Not only that, but it could also increase our safety by enabling devices to detect and alert us to any potential danger in our environment. SIoT is poised to change the world for the better, and its impact on our lives should not be underestimated. This technology will bring about a much-needed revolution in how we interact with our environment and could lead to a more efficient, interconnected future. 

SIoT network of interconnected objects is revolutionizing many industries by making it easier for businesses to track their data more efficiently than ever before. Moreover, IoT devices are now being used in smart homes and factories, allowing users to remotely monitor their surroundings while reducing energy costs by leveraging real-time analytics. Furthermore, they can also provide personalized experiences through voice-activated assistants such as Alexa or Google Home \cite{jones2018voice}. Additionally, they are being leveraged in more creative ways such as developing autonomous vehicles that benefit from high-speed wireless networks \cite{van2018autonomous}. On top of all these advantages, the SIoT takes the potential of IoT a step further. It allows devices to be connected like people -- forming social networks of devices and highly interactive relationships between devices \cite{atzori2012social}. By combining human social networks with the power of IoT, companies can increase customer engagement and loyalty, gain insights into user behavior, create better customer experiences, automate customer service processes, and more. In addition to these benefits for businesses, individuals are also reaping rewards from this technology. The SIoT provide its users with increased safety measures such as automated emergency notifications; improved healthcare opportunities; more efficient transportation; better access to education; enhanced collaboration tools; and improved energy efficiency \cite{trust2vec,vesonet}. These advantages make it easier than ever for individuals to connect with the world around them while using fewer resources and making their lives easier overall \cite{shahab2022siot}.

SIoT environments can be incredibly complex, as they necessarily require a variety of services to be running and available to both users and devices in order to operate \cite{shahab2022siot}. This diversity of services creates challenges when it comes to recommending the right ones for specific groups of users or devices. Therefore, the service recommendation is an important part of SIoT, as it helps to ensure that both users and devices are able to discover appropriate services and applications based on their needs and preferences \cite{obour2019secured}. A service recommendation system utilizes algorithms to collect user data, analyze the collected data, and then suggest services and applications that best suit the users' and devices' profiles. It is important for a service recommendation system to be integrated into the SIoT as without such a system; SIoT entities may become overwhelmed by overwhelming numbers of services and applications available in such a heterogeneous environment \cite{social_computing}. 

Furthermore, integrating a service recommendation system into the SIoT is essential for providing meaningful recommendations tailored to individual user needs due to its heterogeneous nature \cite{ASI}. As well as utilizing sophisticated algorithms which utilize various sources of data, it is important for these systems to maintain an up-to-date database of available services and applications in order to ensure that users are presented with only relevant items when searching. Furthermore, incorporating personalization features will help create engaging experiences tailored to individual user needs more effectively than any manual selection process could ever do. However, due to the SIoT heterogeneous environment, it is becoming increasingly difficult to provide tailored recommended services that best suit user preferences.  Additionally, given that data generated within this environment is often varied across several types and sources it becomes ever more pressing to develop a system capable of making effective recommendations based on multiple modalities. In addition to that, SIoT environment requires multiple services to be running and required by both users/devices, the diversity of the services creates challenges on which service to be recommended to specific groups of users/devices.

Several works have attempted to create general service recommendation systems within the SIoT, however, not all models are well suited for this dynamic environment; particularly when considering different modalities associated with user preference and content heterogeneity. In addition to that, none so far has explored the potential use of multi-modal recommendation which can vastly outmatch single mode. In addition to that, the studies were focusing on users-users and users-items relationships in service recommendation systems. Therefore, it becomes relevant to explore the different modality of generated data in the SIoT to provide tailored service recommendation that suits the needs of SIoT object including users/devices. 

By analyzing the social correlation of service requirements, one potential solution proposed in \cite{kang2017srs} named SRS which is A Social Correlation Group based Recommender System (SRS). SRS generates target groups depending on social correlation of the service requirement using an architecture and procedures based on similar features and principles from Collaborative Filtering and Content-based Recommender Systems. However, the social correlation service recommendation system in SIoT is limited in that it only relies on profile similarity, friend similarity, and interest individuality. This means that it fails to take into account the different modalities of data and the diversity of content which exist across devices and users. Furthermore, this system neglects the ever-changing preferences, interests and needs of users, who may evolve over time. Additionally, the methods used to generate social correlation groups such as Collaborative Filtering and Content-based Recommender Systems are also limited in their capacity to provide accurate recommendations as they are unable to effectively integrate new data into their model and only rely on historical user-item interactions. This means that users may not receive personalized recommendations suitable for their current situation or preferences when there are different modalities of generated data\cite{sun2015recommender}.

To assist users in locating relevant smart objects through the SIoT, a time-aware smart object recommendation model was proposed in \cite{chen2019time} that considers user's preference over time and the social similarity of the objects. This method learns users' preferences from their usage events with a latent probabilistic model, estimates object's social similarities via embedding heterogeneous relationships into shared space, and generates recommendations using collaborative filtering. 

However, similarly, with the previous work, the major drawback of this system is that it does not take into consideration multi-modal data generated in a SIoT environment. It only uses user preference events to learn user's preferences, and estimates objects' social similarities via embedding heterogeneous relationships into shared space from which recommendations are made using collaborative filtering. As such, considering multi-modal data generated within the SIoT environment, as well as contextual factors such as location and user characteristics, should be taken into account to enhance the accuracy and personalization of recommendations delivered by this time-aware smart object recommendation model \cite{contess}.

The works in \cite{chen2021graph} presented a graph-based service recommendation framework was proposed by jointly considering social relationships between heterogeneous objects in SIoT and user's preferences. The User's preferences were learned from their object usage events with a latent variable model while users, objects and their relationships were modeled using a knowledge graph. However, the limitation of taking user's preferences to form a graph knowledge to tailor the recommendation may lead to a lack of accuracy and reliability in providing relevant services from the system. In contrast, due to the lack of consideration for these types of data, the system could potentially overlook important trends or correlations of user-object. Furthermore, this could lead to inaccurate user preference modeling and a decrease in service recommendations that are tailored for individual users. 

Khelloufi et al. \cite{khelloufi2020social} presented a service recommendation system that takes advantage of the different types of relationships between devices' owners when recommending services. In the proposed system, we leveraged a service recommendation system that takes the advantage of device's owners' relationship within specific communities using our effective community detection algorithm to provide tailored service recommendations. However, in this work, the system takes into account the user-user interaction and does not take into consideration the item-item structure in the case of large sparse data.  

In light of this lack of research on multi-modal user preference service recommendation systems in the SIoT, we aim to explore the potential for such systems and provide new insight into their design and implementation. we proposed an advanced multi-modal service recommendation framework specifically tailored for applications in SIoT environments with rich datasets at its disposal.

The main contributions of the proposed framework are as below:

\begin{itemize}

\item  Developed an SIoT service recommendation system that considers the diversity of data generated in the SIoT environment. The proposed system analyses the multi-modal features such as item-item relationships to provide tailored service recommendations in SIoT environment.

\item  Incorporated device heterogeneity as well as data modality into the recommendation process by taking into account different types of devices/data and their capabilities and resources. 

\item  Provided an adaptive service recommendation system that can learn from item-item structure and improve the accuracy of future recommendations. 
\end{itemize}

Extensive experiments were conducted to validate the performance of the proposed service recommendation system using relevant and well-known performance metrics, based on real-world datasets. The results demonstrate the effectiveness of the system in providing accurate recommendations.

The rest of this paper is organized as follows: Section \ref{sec.Relatedworks}  reviews the recent literature on service recommendations in SIoT. In Section \ref{sec.SystemDesign}, we present the system modeling, where we discuss the SIoT object-object relationships, object-service events, and the KNN modality-aware graph. The experimental evaluation is discussed in Section \ref{sec.Evaluation}. Finally, we conclude the paper and outline future research directions in Section \ref{sec.Conclusion}.

\section{Related Works}
\label{sec.Relatedworks}

The Social Internet of Things (SIoT) has introduced new opportunities for providing personalized and context-aware service recommendations to users. As the SIoT ecosystem expands, various types of service recommendation systems have been developed to cater to the diverse needs and preferences of users in this socially connected environment. This section presents an overview of different service recommendation system types specifically tailored for the SIoT, highlighting their unique characteristics and approaches. By exploring these distinct recommendation types, we aim to provide insights into the advancements and potential applications of service recommendation systems in the SIoT domain. From collaborative filtering and content-based recommendation to hybrid and social-based approaches, each system type offers its own set of advantages and challenges in enhancing user experiences and facilitating effective service discovery in the SIoT realm.

The framework presented in \cite{chen2021graph} adopts a graph-based approach, considering both the social relationships among heterogeneous objects in the SIoT and user preferences. User preferences are learned from object usage events using a latent variable model, while the relationships between users, objects, and services are modeled using a knowledge graph. The service recommendation task is treated as a knowledge graph completion problem, aiming to predict the "like" property that connects users to services. The researchers have built a SIoT testbed to validate the proposed model, and experimental results demonstrate its feasibility and effectiveness in enhancing service recommendation within the SIoT environment. 

Authors in \cite{chen2019time}  proposes a time-aware smart object recommendation model that combines user preferences over time and the social similarity of smart objects. The model leverages a latent probabilistic approach to learn user preferences and embeds heterogeneous social relationships of smart objects into a shared lower-dimensional space to estimate social similarity. The recommendation list is then generated using item-based collaborative filtering. The proposed method is evaluated on two real-world datasets, demonstrating superior recommendation effectiveness compared to baseline approaches. However, this model relies on object usage events to learn user preferences, potentially limiting its effectiveness if data on object usage is incomplete or unavailable for certain users. However, this approach overlooks crucial factors like user context, preferences, and specific requirements, focusing solely on the "like" property to connect users with services. Consequently, recommendations may lack personalization and accuracy. 

The study presented on \cite{zhang2021mining}  focuses on designing a recommender system for the SIoT. The recommender system leverages the social dynamics influencing the behavior and interactions of autonomous objects within the SIoT. The primary objective is to facilitate optimal pairing of objects to enable effective recommendations, aiming to provide the best possible results to users. The described recommender system for the SIoT heavily relies on social dynamics, which may overlook other important factors such as user preferences and contextual information, potentially leading to less personalized and limited recommendation accuracy.

In our previous work we proposed a service recommendation system that leverages the social relationships between device owners to improve the accuracy and diversity of offered services in the IoT context \cite{khelloufi2020social}. However, the presented service recommendation system we overlooks the potential limitations of relying solely on users' social relationships, neglecting other crucial factors such as user preferences, context, and specific requirements. Consideration of these additional aspects could further enhance the efficiency and personalization of the recommended services in the IoT environment.

Authors in \cite{chen2021graph} presented a graph-based service recommendation framework for addressing service recommendation challenges in the context of SIoT. The framework utilizes a knowledge graph representation, where users and SIoT objects are treated as entities. The objective is to predict the implicit "like" property that connects users to specific services based on their preferences. Object usage events, recorded through RFID and sensor readings, provide rich information for uncovering user preferences. The framework decomposes SIoT service recommendation into a link prediction process that considers user preferences and object social relationships. However, the scalability of the proposed graph-based framework might be a concern. As the number of users, objects, and services increases, the size of the knowledge graph can grow significantly, which could impact the efficiency and computational requirements of the recommendation process. In addition to that, the framework does not deal with the cold start problem, as it faces difficulties when dealing with new users or objects that have limited or no usage history. Since the recommendation relies on user preferences and object usage events, it may struggle to provide accurate recommendations for such cases. 

However, since user ratings or preferences for SIoT services are often implicit, traditional recommendation models like content-based filtering and collaborative filtering may not be suitable. In the SIoT context, data is diverse and can come in various modalities such as images, videos, audio, and text. This diversity presents a unique challenge for service recommendation, as each modality can provide valuable information about user preferences and service characteristics. To address this challenge, there is a growing need for multi-modal service recommendation approaches that can effectively integrate and leverage information from different modalities. By considering multiple modalities, such as analyzing images to understand user context or utilizing text data to capture user preferences, we can gain a more comprehensive understanding of user needs and provide more accurate and personalized recommendations in the SIoT environment. The development of multi-modal service recommendation techniques holds great promise for improving the quality and relevance of recommendations in the complex and dynamic world of SIoT.

\section{System Design}
\label{sec.SystemDesign}
With the rise of SIoT, the number of connected devices has increased significantly, leading to an overwhelming amount of data being generated. As a result, users face difficulties in finding the most suitable service or product among the vast options available. In this scenario, a personalized service recommendation system is crucial to provide users with personalized and relevant recommendations based on their preferences, usage patterns, and social network information. The system should be able to learn from user interactions and dynamically adapt to their evolving needs, thus, improving the user experience and customer satisfaction. The design of the proposed multi-modal service recommendation system in the SIoT involves several key components that enable personalized and context-aware recommendations. The system aims to leverage the social interactions and data generated by interconnected IoT devices to provide users with relevant and valuable services. In this section, we present an overview of the system architecture and the main components involved.

\begin{figure}
\centering
\includegraphics*[width=3.39in, height=2.86in]{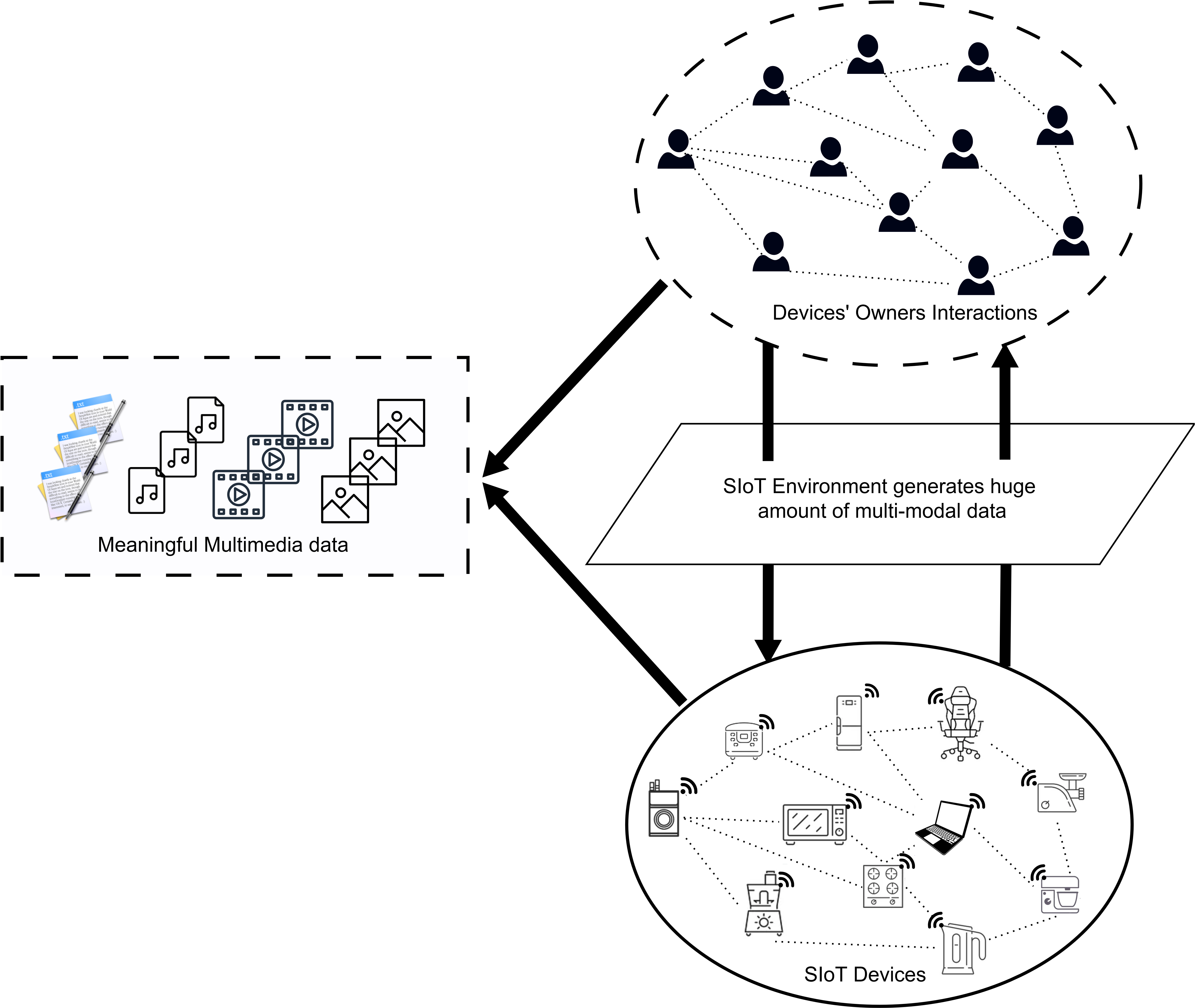}
\caption{multi-modal generated data in SIoT environment.}   
\label{siot_envi}
\end{figure}

Figure \ref{siot_envi} illustrates the generation of a vast amount of diverse and multi-modal data by devices in the SIoT ecosystem. The types of data include images, videos, textual content, audio, and more. This data is produced by interconnected devices that are part of the SIoT, such as smartphones, wearables, smart home devices, and environmental sensors. Therefore, the importance of developing a sophisticated service recommendation system in response to this data explosion can not be overstated. In the subsequent subsections, we delve into the key components and system design of the proposed service recommendation system for SIoT. First, We delve into the details of system modeling where we outline the architectural design and data flow of our recommendation framework. Next, we explore the used technique of item-item latent structure learning, which enables us to capture the underlying relationships and dependencies between diverse data items. We then introduce the innovative KNN Modality-Aware Graph, a novel approach that leverages multi-modal data to enhance recommendation accuracy. Additionally, we delve into the methodology of multi-modal latent graph aggregation, which consolidates information from various modalities to generate comprehensive and context-aware recommendations. Finally, we culminate with a compelling scenario case study that demonstrates the practical application of our recommendation system in a multi-modal data-rich SIoT environment. Through these subsections, we present a comprehensive overview of our advanced service recommendation system and its efficacy in navigating the complexities of the SIoT landscape.

\subsection{System modeling}
In this section, we begin by formally defining the main data structures and notations utilized in our system. This serves as a foundation for understanding the subsequent discussions. Additionally, we outline the problem definition for service recommendation within the context of the SIoT. We then delve into the details of the item-item latent structure learning process, the KNN modality aware graph, and finally, we present and discuss our proposed multi-modal graph aggregation method. These components play a crucial role in enhancing the accuracy and effectiveness of our service recommendation system within the SIoT framework.


In the SIoT system, each SIoT object $O_i$ is represented by a tuple $<S^j_i,\ U^k_i>$ , where $S^j_i$ is the set of services provided by this device \textit{i}, and $U^k_i$are the set of owners use/own this device. Additionally, the set of services in the system is denoted by:
\[S=\{s_1,s_2,s_3,\dots \dots s_n\}. \] 
The SIoT relationship graph denoted by  $G=\left\{O,\ R,\ U\right\}$  consists of three components:

$O$ represents all objects in the network, 

$R$ represents the relationships between these objects, 

And  $U$ represents users that own specific devices.


The Object-Service Event Set is the combination of two separate events: Formally known as \textbf{ \textit{object-service usage event}} and \textbf{ \textit{object-service generation event }}which are denoted by $\textbf{UE}$ and $\textbf{GE}$ respectively in timestamp set \textit{TS}.

\noindent An object-service usage event set defined by UE is the set of elements ${ue}^{(ij)}$ 
\begin{equation}
 \label{Eq.1}
UE=\left\{{ue}^{(ij)}_1,{ue}^{(ij)}_2,{ue}^{(ij)}_3\dots .{ue}^{(ij)}_m\right\}   
\end{equation}
Where,  ${ue}^{(ij)}$  is a single object-service usage event.

\begin{equation}
\label{Eq.2}
\begin{aligned}
&      {ue}^{(ij)}=S^{(j,\delta t)}_i        \\ 
&       S^{(j,\delta t)}_i=  \\
&{<o_i,\ s_j,u_k,\delta t>|o_i\in O\ \wedge s_j\in S^j_i\wedge u_k\in U^k_i\wedge \delta t\in TS}
\end{aligned}
\end{equation}

i.e. a service $S^j_i\ $ is used by an object $O_i\ $in a timestamp  $\Delta t$=$t_s-\ t_e$

An object-service generation event set defined by GE is the set of elements ${ge}^{(ij)}$ 

\begin{equation}
 \label{Eq.3}
GE=\left\{{ge}^{(ij)}_1,{ge}^{(ij)}_2,{ge}^{(ij)}_3\dots .{ge}^{(ij)}_{m'}\right\}    
\end{equation}
Where,  ${ge}^{(ij)}$  is a single object-service generation event.

\begin{align}
\label{Eq.4}
    \begin{array}{l}
      {ge}^{(ij)}=S^{'(j,{\Delta t}')}_i       \\ 
       S^{'(j,{\Delta t}')}_i= \\ 
       <o_i,\ {s_j}',u_k,{\Delta t}'>|\ \ \\
       o_i\in O\ \wedge {s_j}'  \in S^j_i\wedge u_k\in U^k_i\wedge {\Delta t}'\in TS
    \end{array}
\end{align}
i.e. a service $S^{'j}_i\ $ is generated by an object $O_i\ $in a timestamp  ${\Delta t}'$=$t'_s-\ t'_e$

From equations \ref{Eq.2} and \ref{Eq.4}  the SIoT object service event indicates that a single user $u_k$ has used and generated 
two services $s_j$ and ${s_j}^\prime$ in a timeframe $\Delta t$ and $\Delta t^\prime$ respectively.

A service explosion problem is defined as a situation in which a huge amount of services/data are both produced and consumed by a set of devices in the system: 

\begin{align}  
\label{Eq.5}
\begin{array}{l}
      S_{exp}=\left[UE.GE\right] . TS  \\ 
      =   \bigcup_{ _{l=1..\left|UE\right| 
           l'=1..\left|GE\right|} } \\ 
           {\left({ue}^{\left(ij\right)}_l,{ge}^{\left(ij\right)}_{l'},\Delta t\right)}
    \end{array}   
\end{align}

SIoT Object-Service Events defined by the two object-service usage events $UE$ and object-service generation event $GE$  in service explosion situations $S_{exp}$ create challenges when it comes to recommending the right services for specific groups of users or devices. Therefore, the service recommendation is an important part of the SIoT, as it helps to ensure that both users and devices are able to discover appropriate services and applications based on their needs and preferences

\subsection{Item-item Latent structure learning}
In the SIoT environment, the multi-modal items provides valuable features.  Existing methods only use these multi-modal items as side information neglecting the important semantic relationships between underlying features and item-item relationships, in this section, we specifically focus on uncovering latent graph structures of item graphs to learn more accurate representations. To do so, inspired by \cite{ashton2009internet} the modality-aware k-nearest-neighbor (kNN) item graph is first built using raw multi-modal data in order to develop modified and advanced feature extraction techniques. Finally, these individual latent graphs are combined together into one unified structure with a consistent method for analysis.

\begin{figure}
\centering
\noindent \includegraphics*[width=3.53in, height=3.15in]{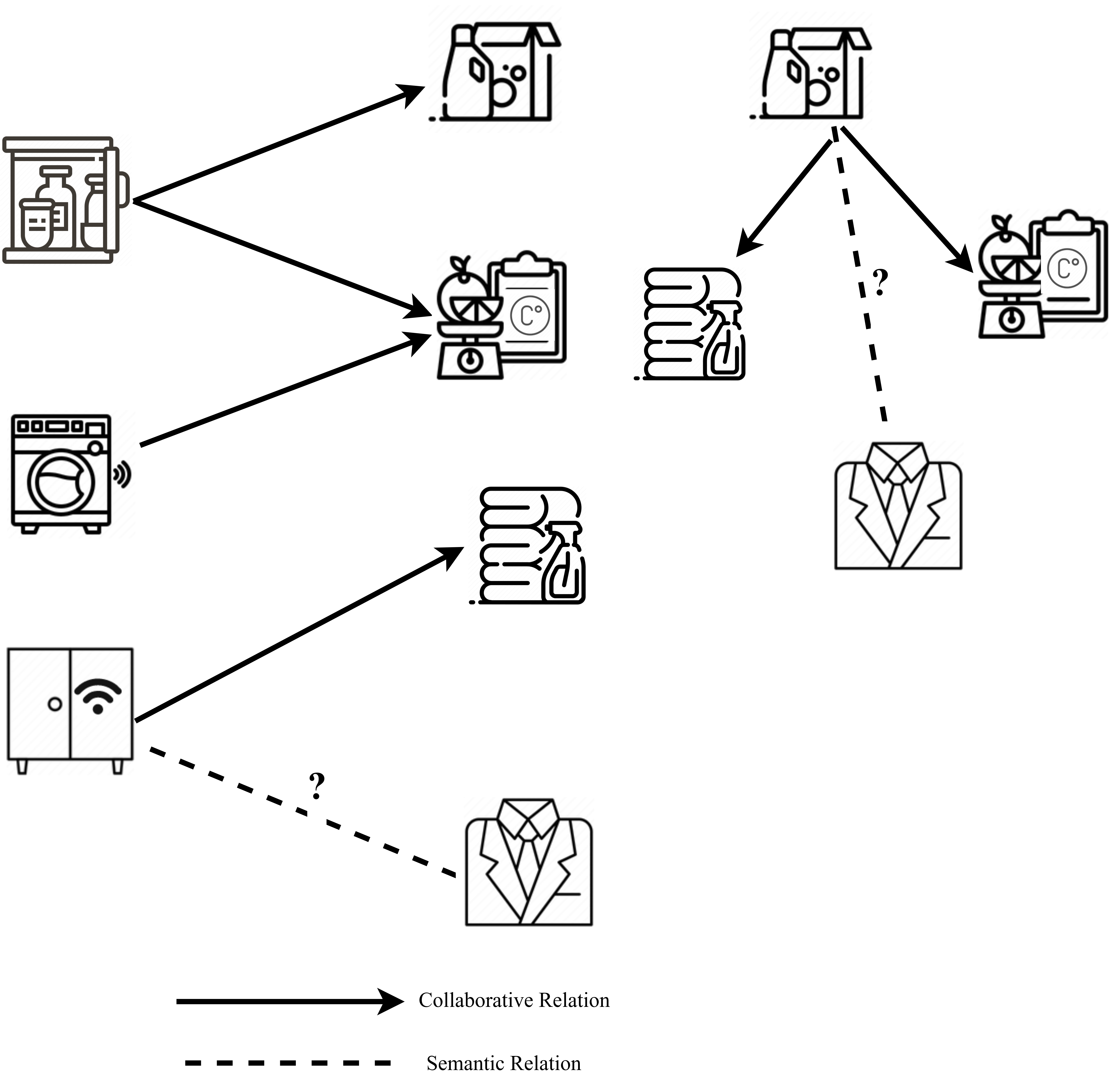}
\caption{Item-item learning structure of data exchanged between SIoT devices }
\label{image_2}
\end{figure}

This approach goes beyond standard methods by providing a deeper understanding of items' semantic relationships through their underlying qualities. By using this approach to create and analyses modality-aware k-nearest-neighbor item graphs, it is possible to access richer, more meaningful content information than what can be obtained through traditional methods that rely solely on multi-modal features as side information for each item. Through transforming raw multi-modal features into higher-level representations and identifying the key semantic relationships among them via learning latent graph structures, it is possible to construct more precise item depictions and enable cross-modal integration in a versatile manner. Figure \ref{image_2} illustrates the structure of item-item learning, showcasing the collaborative relationship between devices and services. It highlights that within this relationship, there exist hidden latent structures that can be explored to uncover semantic item-item connections. These connections enable a deeper understanding of the underlying associations and interdependencies between different items, leading to more effective and contextually relevant recommendations in the recommendation system.

\subsection{k-Nearest-Neighbor (KNN) Modality-Aware Graph}

KNN modality-aware graph construction is a form of data processing which involves creating a graph based on the initial kNN modality-aware graph kk. This graph is created using raw features from each modality, such as audio, visual and textual classifiers. The hypothesis behind this approach is that similar items are more likely to interact than dissimilar items, so it is important to quantify the semantic relationship between two items through similarity measures. Commonly used methods include cosine similarity \cite{al2020internet}, kernel-based functions \cite{jones2018voice} and attention mechanisms \cite{van2018autonomous}. Once the similarities have been calculated, an adjacency matrix can be created which represents the relationships between nodes in the graph. Generally speaking, these graphs are much sparser than fully connected graphs as they tend not to contain noisy or unimportant edges \cite{atzori2012social}. Therefore, any entries with negative values will be suppressed to zero and thus, the modality matrix of the features $M^f \in {R}^{N \times N}$ is calculated by :

\begin{equation}
 \label{Eq.6}
M^f_{ij}=\frac{{\left({\textbf{e}}^f_i\right)}^{\mathrm{T}}{\textbf{e}}^f_j}{\left\|{\textbf{e}}^f_i\right\|\left\|{\textbf{e}}^f_j\right\|}    
\end{equation}

\begin{equation}
 \label{Eq.7}
\widehat{M^f_{ij}}=\left\{ \begin{array}{c}
 \begin{array}{c}
M^f_{ij}\ \ \ \ \ \ \ M^f_{ij}\in top-k\left(M^f_{ij}\right) \\ 
 \\ 
0\ \ \ \ \ \ \ \  otherwise, \end{array}
\ \ \ \  \\ 
\ \  \end{array}
\right\}    
\end{equation}

The matrix $\widehat{M^f_{ij}}$ is the generated directed graph adjacency matrix. The exploding or vanishing gradient problem occurs when gradients of the errors in deep networks become exponentially larger or smaller \cite{pascanu2013difficulty}, respectively. to prevent this issue, the Normalization of the adjacency matrix is applied as below:

\begin{equation}
 \label{Eq.8}
{\tilde{M}}_f={\left(D^f\right)}^{-\frac{1}{2}}{\hat{M}}^f{\left(D^f\right)}^{-\frac{1}{2}}   
\end{equation}
\begin{equation}
 \label{Eq.9} 
D^f_{i,j}=\left\{ \begin{array}{c}
{\mathrm{deg} \left(v_i\right)\ }\ \ \ \ \ if\ i=j \\ 
 \\ 
0\ \ \ \ \ \ \ \ \ \ otherwise \end{array}
\right\}
\end{equation}
Where:
\noindent $D^f\in {\mathbb{R}}^{N\times N}$is the diagonal degree matrix of the matrix ${\hat{M}}^f$, and

\begin{equation}
 \label{Eq.10}
D^f_{ij}=\sum^{k\hat{M}}_{i,j}{{\hat{M}}^f_{ij}}\   
\end{equation}
The obtained modality-aware initial graph structures $\widetilde{M^f}$via raw multi-modal features can be beneficial to the recommendation task. However, these initial graphs may not effectively reflect underlying graph structures due to noise and incomplete data measurement or collection. To shift this paradigm, we present a method of learning dynamic graph structures from transformed high-level multi-modal features and integrating them with existing ones for successful recommendations. 

To capture the underlying relationships within high-level feature vector $\widetilde{h^f_i}$, we dynamically infer graph structures by utilizing a trainable transformation matrix $T_f$ and bias vector $b_f$.  Therefore, the high- level multi-modal feature $\widetilde{H^f_i}$ is obtained as below:

\begin{equation}
 \label{Eq.11}
{\tilde{h}}^f_i=T_fh^f_i+b_f\   
\end{equation}

The graph learning process in equations (6, 7 and 8) is repeated until we obtain the adjacency matrix ${\tilde{A}}^f$

The initial item graph may contain a large amount of superfluous information, yet it is still rich in valuable structural data. To ensure this knowledge can be incorporated into the learning process and also minimize training instability due to drastic variations in adjacency matrices, we introduce skip connections $\sigma \in (0,1)$ which combine learned graphs with their original structure. Therefore, the final graph normalized and sparsified adjacency matrix is represented as below:

\begin{equation}
 \label{Eq.12}
A^f=\sigma {\tilde{M}}_f+\left(1-\sigma \right){\tilde{A}}^f\  
\end{equation}

\subsection{Multi-modal latent graph aggregation}

In this section, we describe how we merge the different modalities to obtain the final latent structure. In an SIoT multi-modal recommendation system, devices/users may focus on specific modalities to provide tailored recommendation. For instance, when a device owner is shopping for clothes, they may want to use the SIoT device to access recommendations from other devices whom owners have tried on similar items. In this case, the user would likely be more focused on visual modalities such as images or videos of how the items look on different people. The device would provide images and videos with accompanying text descriptions so that the user can get a better idea of what they are purchasing. On the other hand, when selecting books, the user will likely focus more on textual information such as reviews from other readers or blurbs about the book's content. In this case, the SIoT device would provide access to reviews written by experienced readers and descriptions provided by booksellers. Therefore, an adaptive weight to balance modality-specific graphs is required as an importance score which is presented as below:

\begin{equation}
 \label{Eq.13}
A=\sum^{|F|}_{f=0}{{\alpha }_mA^f}\  
\end{equation}
${\alpha }_m$ is the score given for each modality
To ensure that each modality graph is normalized, we apply a SoftMax activation function \cite{sharma2017activation} in computing the final adjacency matrix $A$ where ${\alpha }_m$ represents an importance score assigned to all modalities. This allows us to obtain accurate and meaningful latent structures while leveraging multiple sources of user data for enhanced recommendations. Subsequently, we apply graph convolution operations to generate enhanced item representations by including item-item similarities in our embedding approach. Similar to messages being transmitted and merged together, this process allows for one item's representation to be enriched through gleaning information from its neighbor peers. By stacking multiple layers of these graph convolutions, high order relationships between items can also be identified and incorporated emphasizing a more complete picture than prior techniques would allow for. Inspired  by works in \cite{he2020lightgcn} , we used a simple message passing with aggregation without need for any additional feature transformations or activations delivers strong results that are both effective and computationally efficient . The k-th layer is characterized as follows:

\begin{equation}
 \label{Eq.14}
L^{(l)}_i=\sum^n_{item\in N(i)}{A_{i,item}L^{(l-1)}_{item}}\ 
\end{equation}

\subsection{Optimization and combining collaborative filtering }

Our proposed approach trains the model to learn item-item representations using multiple features and integrates them with downstream collaborative filtering methods that simulate user-item interactions. This method is adaptable and can be utilized as a module with any collaborative filtering approach. The resulting user and item embeddings are denoted as \textbf{\textit{xu}} and \textbf{\textit{xi}}, respectively. The item embeddings are boosted by adding normalized embeddings hl obtained from the item graph.

\begin{equation}
 \label{Eq.15}
{\hat{O}}_i={\tilde{O}}_i+\frac{L^{(l)}_i}{{\left\|L^{(l)}_i\right\|}_2}\
\end{equation}

\begin{equation}
 \label{Eq.16}
{\hat{X}}_{u,i}=<{\tilde{O}}^{\mathrm{\Pi }}_u,{\hat{O}}^T_i>\
\end{equation}

Subsequently,  the Bayesian Personalized Ranking loss is utilized to calculate pairwise ranking, promoting the estimation of the observed entry over its unobserved counterparts.

\begin{equation}
 \label{Eq.17}
\gamma_{BPR}= -\sum_{u\in U}\sum_{i\in I_u}{\sum_{j\notin I_u}{{ln}_{\varsigma}({\hat{X}}_{u,i}-{\hat{X}}_{u,j})}}
\end{equation}

Where $I_u$ represents the item linked with a user u, the pairwise training tiple is denoted by (u, i, j) and the sigmoid function is represented by $\varsigma$

\begin{figure}
\centering
\includegraphics*[width=3.54in, height=3.80in, trim=0.00in 0.00in 0.00in 1.21in]{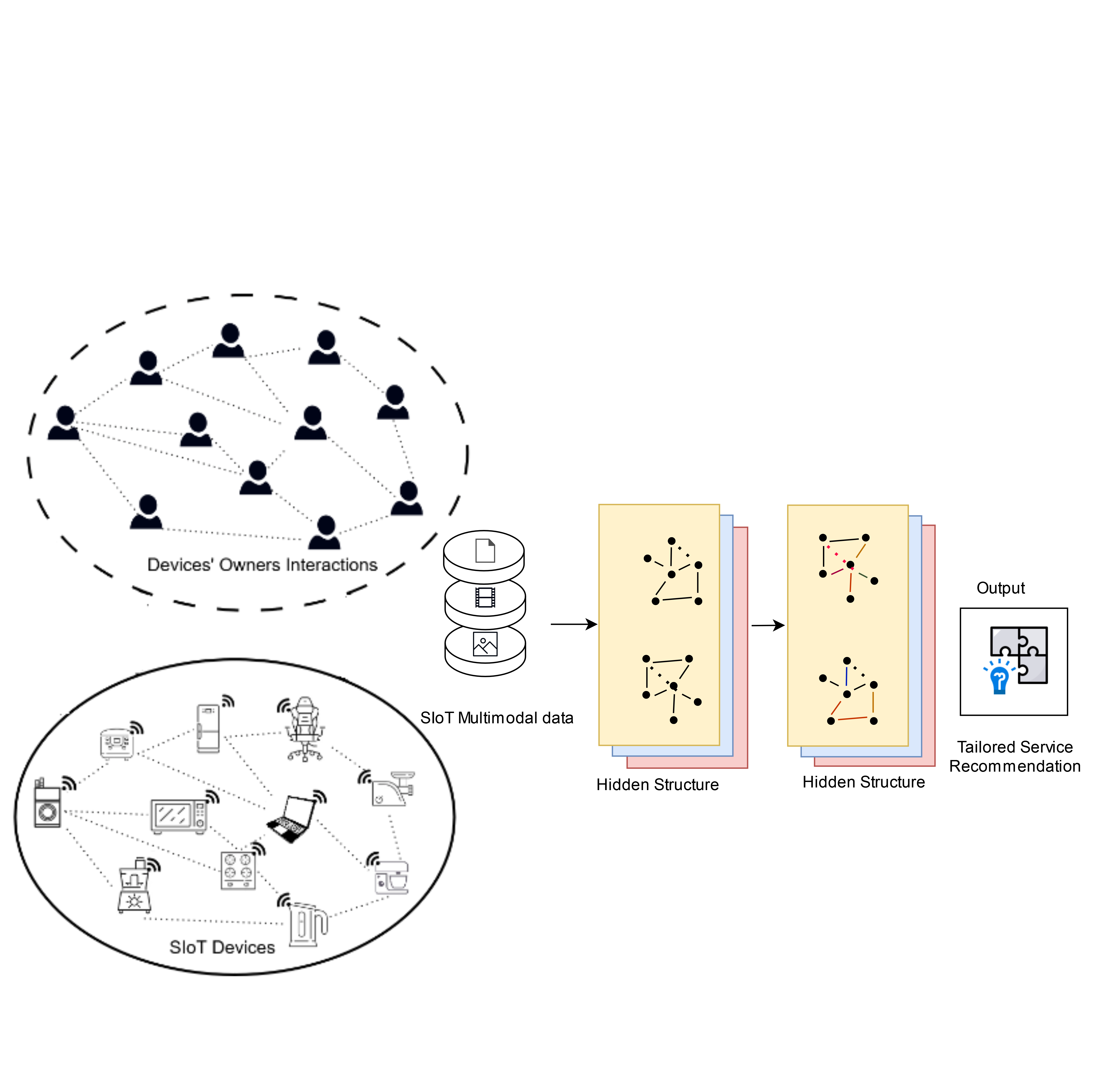}
\caption{Latent structure exploration for tailored recommendation in SIoT environment}
\label{image_3}
\end{figure}

In summary,  Figure \ref{image_3}  illustrates the key structure of the SIoT environment, where both users and devices contribute to shaping the SIoT. Users have their own social relationships, which are also mapped to their respective devices. In this paper, we argue that the vast amount of data generated by this interconnected network holds hidden structures that can be represented as structural graphs. These graphs enable tailored service recommendations, making them essential in the field of service recommendation.

\begin{figure*}
\centering
 \noindent \includegraphics*[width=5.52in, height=3.56in]{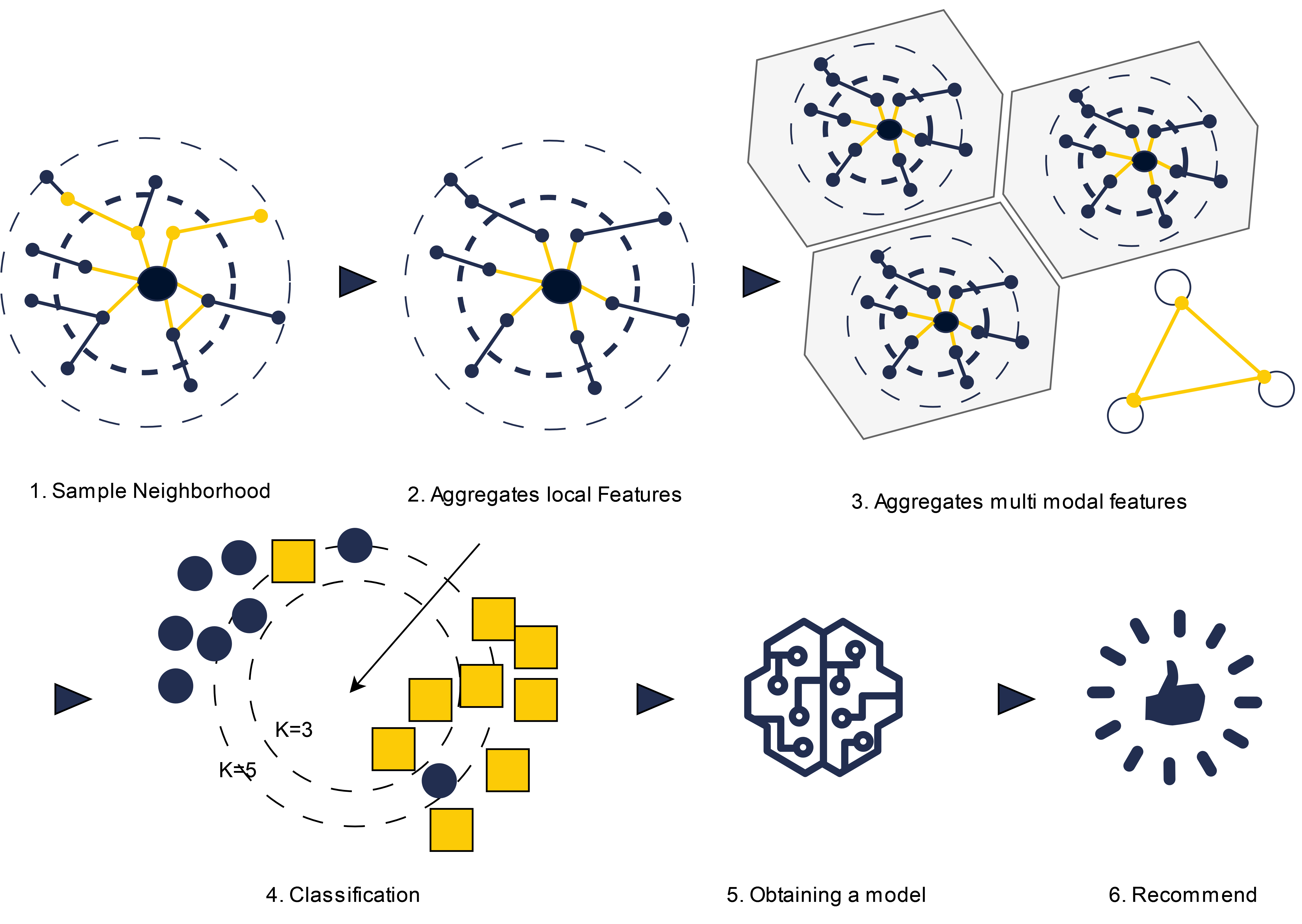}
  \caption{Recommendation Process.}   
\label{image_4}
\end{figure*}

On the other hand, Figure \ref{image_4} depicts the sequential steps of the recommendation process. Initially, a neighborhood of similar items or users is sampled based on similarity measures. Then, the system aggregates local features by extracting relevant information from each item or user within the neighborhood, including ratings, preferences, and historical behavior. Additionally, multi-modal features associated with items or users, such as images, text, audio, or video, are integrated and processed for a comprehensive representation. The K-Nearest Neighbors (KNN) classification is applied to classify items or users based on their local and multi-modal features, identifying the most similar ones. The system obtains a model that captures relationships and patterns, serving as the foundation for generating personalized recommendations. These recommendations consider similarities, local features, and multi-modal information to provide relevant suggestions or predictions. Overall, this approach leverages neighborhood sampling, feature aggregation, KNN classification, and model creation to enhance the accuracy and relevance of recommendations in the recommendation generation process.  

\subsection{Scenario cases of latent structure service recommendation in SIoT}

This section presents four distinct scenarios in the SIoT environment that provide insights into the importance of multi-modal service recommendations for various domains. The scenarios include a factory production line, an intelligent home security system, a smart city monitoring platform, and a healthcare organization. In each scenario, different SIoT devices and sensors collect data, enabling real-time monitoring and analysis. Through the extraction of latent structures from the collected data, these systems can provide tailored recommendations, ranging from device upgrades and replacements for enhanced efficiency, to improved security measures, optimized city infrastructure, and personalized healthcare services. These scenarios highlight the potential of leveraging SIoT and data-driven insights to optimize operations and improve outcomes in diverse domains.

\subsubsection{Industrial SIoT}
A factory production line specializing in manufacturing a specific product is connected to an array of SIoT sensors and devices, which can monitor the manufacturing process in real time while they created different SIoT relationships. To improve efficiency, the machines on the production line are able to recommend service upgrades and replacements for each device in order to ensure maximum performance. Through the different factors that take into account usage metrics such as energy consumption, temperature, and vibration levels, the system will be able to determine when a device should be serviced or replaced before it fails. This proactive approach helps prevent unexpected breakdowns or disruptions in the production line, minimizing downtime and optimizing productivity. This scenario case recommendation uses textual features from raw text collected data.

\subsubsection{SIoT security system}
An intelligent home security system can monitor its environment and detect potential threats or suspicious activity. The system has access to a range of data points including motion detection, audio recordings, and video recordings from different SIoT devices such as cameras and other sensors. The system can then make recommendations for services and upgrades based on the data it collects. For instance, if the system detects suspicious activity, it might provide recommendations for adding additional cameras or upgrading existing ones for improved coverage and protection of the property. It could also suggest the installation of more secure locks with better encryption or strong authentication systems as well as additional lighting around potential entry points such as windows or doors. Such a system needs the latent visual structure from the collected data to efficiently manage tailored recommendation.  

\subsubsection{SIoT for smart city}
A smart city monitoring platform has been developed with access to a wide range of data sources including traffic cameras, air quality sensors and weather stations around the city. This multi-modal data is analyzed and latent structure can be extracted to predict congestion levels at different times of day as well as forecast future events such as heavy rain or extreme temperatures that may affect infrastructure performance or public safety measures such as air pollution levels in certain areas. Based on this analysis the platform can recommend specific services such as road maintenance works or upgrades to traffic lights that will help reduce congestion and improve public safety conditions in particular areas of the city. Similarly, the collected data varies from different types and thus the need for latent structure is mandatory to provide suitable recommendations.

\subsubsection{SIoT-enabled Healthcare}
A healthcare organization has deployed an IoT platform to monitor patient health across different locations within its network of clinics and hospitals. Different sensor devices are used to collect vital information from patients such as their heart rate, blood pressure and temperature readings which can then be monitored remotely by healthcare professionals in real-time through cloud-based analytics platforms powered by artificial intelligence (AI). Based on this data collected from patients over time the platform can recommend specific services such as lifestyle changes for particular individuals based on their individual health profile or treatments tailored specifically for them according to their medical history which may have been accumulated over multiple visits with different doctors within the organization's network.

\section{Experimental Evaluations}
\label{sec.Evaluation}
This section covers several important aspects of our research, including the research questions, the used dataset, baselines, and performance evaluation metrics.

We used three comprehensive Amazon datasets that encompass a wide range of products and user interactions. These datasets offer a diverse collection of user-item interactions, ratings, and multi-modal contextual information, providing a rich foundation for our experiments. To establish a performance baseline, we selected a state-of-the-art service recommendation model specifically designed for the SIoT, which is up-to-date in the literature. Additionally, we defined a set of evaluation metrics that encompass various aspects, such as precision, recall, and mean average precision. These metrics enable a comprehensive assessment of the effectiveness and accuracy of our proposed recommendation system. 

Our experimental settings are designed to address the research questions posed in this study. Firstly, we conducted experiments to compare our model against current service recommendation techniques in SIoT, both in warm-start and cold-start settings. We evaluated the predictive performance of each method using metrics such as precision, recall, RMSE, and MAE. Furthermore, we explored the utility of the item graphs generated from multi-modal features in the context of SIoT service recommendation. By extracting embedding features from each dataset and utilizing them to construct an item graph structure, we investigated the effectiveness of these structures in our recommendation task. Lastly, we performed sensitivity analysis to assess the impact of key hyperparameters, such as learning rate and batch size, on our model's performance. By varying these parameters and observing the outcomes, we gained insights into the sensitivity and robustness of our proposed model.

The research questions addressed in this study are as follows: 

RQ1: How does our proposed service recommendation model compare to existing techniques in SIoT, considering both warm-start and cold-start scenarios? 

RQ2: What is the effectiveness of the item graphs generated from multi-modal features in SIoT service recommendation? 

RQ3: How does altering specific hyperparameter settings impact the performance and outcomes of our proposed model?

Through these investigations, we aim to provide valuable insights into the comparative performance of different recommendation approaches, the impact of multi-modal features, and the sensitivity of our model to key hyperparameters.

\subsection{Dataset and Experimental settings}

In this subsection, we outline the specific steps taken to prepare the dataset for our experiments. To ensure a robust evaluation of our recommendation models, we employed a train-test split approach. The dataset was divided into training and testing sets, with 80\% of the data allocated for training and the remaining 20\% set aside for evaluation. This division allows us to train the models on a substantial amount of data while reserving a separate portion for assessing their performance. 

Due to the lack of a multi-modal dataset for SIoT, we have conducted our experiments on a widely and publicly used Amazon dataset presented in \cite{ni2019justifying}. The dataset includes both visual and textual information in three categories: Clothing, Sports and Baby. The provided dataset has been compiled for research purposes by Amazon Labs. The dataset encompasses a range of distinctive features, including but not limited to:

\begin{itemize}
\item  User/item interactions

\item  Star ratings

\item  Timestamps

\item  Product reviews

\item  Social networks

\item  Item-to-item relationships (such as copurchases and compatibility)

\item  Product images

\item  Price, brand, and category information

\item  GPS data

\item  Heart-rate sequences

\item  Other metadata
\end{itemize}

Consequently, this dataset can be categorized as a multi-modal dataset due to the diverse array of modalities it incorporates.

Table \ref{tab:my_table} shows the summary of the given dataset. For our research, we use the 4,096-dimensional visual features that have been extracted and published. Additionally, in order to incorporate the textual modality, we extract 1,024-dimensional sentence embeddings by concatenating the title, description, categories, and brand of each item and utilizing sentence-transformers.

\begin{table}
    \centering
    \caption{Dataset Description}
\begin{tabular}{ cl cl cl cl cl cl} \hline 
  & Users & Items & Interactions & Sparsity \\\hline   
Clothing & 54,5255 & 25,564 & 25,4756 & 99.88\% \\ 
Sport & 40,254 & 14,235 & 26,365 & 99.85\% \\ 
Baby & 25,654 & 8,542 & 145,523 & 99.70\% \\ \hline 
\end{tabular}
    
    \label{tab:my_table}
\end{table}

\subsection{Baseline}

In order to evaluate the performance of the proposed system, we have established several baselines systems and provide a comparison with four up-to-date approaches including ORTJ  \cite{zhang2022smart} , BLA \cite{zhang2021mining} ,GBSR \cite{chen2021graph} , and MF \cite{koren2009matrix} :

\begin{itemize}
\item  ORTJ \cite{zhang2022smart}: is an object recommendation approach that is proposed to consider both the attribute and text-topic information related to smart objects. Thing thing relationship is introduced as an attribute of smart objects in the IoT to improve recommendation accuracy. An ORTJ model based on maximum a posteriori estimation is derived and experiments are conducted to compare its performance with others.  

\item  BLA \cite{zhang2022smart}: It is a neural network model that has been proposed for the recommendation of smart objects in SIoT based on user requirements. The model utilizes BERT and Bi-LSTM to generate vectors for matching and recommendations. Self and global attention mechanisms are used to dynamically reweight vectors for improved performance. Moreover, Thing--thing relationship data was introduced into the model to utilize user requirements for generating more reasonable representations of smart objects attributes and characteristics. The authors used an extended original MovieLens dataset in SIoT scenarios.

\item GBSR \cite{chen2021graph}: It is a framework that has been presented in order to address the service recommendation problem in SIoT. In this model, the SIoT service recommendation problem is modelled as a knowledge graph completion problem. The user's preference and object usage events with rich spatial-temporal information are used to uncover user's preference. More exactly, the hidden factor of the user's object usage is modeled and knowledge graph is constructed through service usage value which is reflected by service usage frequency.  

\item  Matrix factorization(MF) \cite{koren2009matrix} : It is a collaborative filtering method that is used to improve recommendation accuracy by reducing data sparsity of the user-item matrix. User and item matrices are created, each with a vector for each row/column respectively and the predictive score of user i for item j is calculated by multiplying corresponding vectors in the two matrices. Matrices are adjusted During training to reduce the least squared error between actual and predicted values.
\end{itemize}

In addition to the aforementioned baselines, we also evaluated our proposed solution in terms of cold-start performance with the solution proposed in \cite{yao2016things}. This evaluation was conducted because the previous baselines \cite{zhang2022smart, zhang2023smart} lacked assessment specifically addressing the cold-start problem.

\subsection{Metrics}

To ensure a fair and standardized evaluation, we have adopted the evaluation settings described in prior research [18][19][11]. Our evaluation focuses on top-K recommendation performance and employs various widely-accepted evaluation metrics including: recall, precision, MAE, RMSE, NDCG@K. To provide comprehensive results, we calculate the average metrics for all users in the test set under three scenarios: K = 5, K = 10 and K = 15.

To prepare the data for evaluation, we follow a specific data partitioning scheme. For each user in the dataset under evaluation, we randomly allocate 80\% of their historical interactions for training purposes. Additionally, 10\% of the interactions are reserved for validation, while the remaining 10\% are used for testing. This partitioning allows us to create distinct subsets that capture different aspects of user behavior.

During the training phase, we incorporate a negative sampling strategy. This strategy involves pairing each observed user-item interaction in the training set with a negative item that the user has not previously interacted with. This approach enables us to train the model to distinguish between positive and negative user-item interactions effectively.

For evaluating recommendation accuracy, we employ the all-ranking protocol. This protocol computes the evaluation metrics based on the ranked list of recommended items for each user. By employing this protocol, we can accurately assess the performance of our recommendation models in terms of accuracy and effectiveness.

In this subsection, we will discuss four of the most common baseline metrics used in the experiment: MAE, RMSE, Precision, Recall and NDCG. 

\begin{itemize}
\item  Mean Absolute Error (MAE) is a baseline metric used to measure the accuracy of predictions in predictive modeling. It is an average difference between the predicted and observed values, and is often expressed as a percentage. 

\begin{equation}
\label{Eq.18}
MAE=\frac{\sum_{i,j}{|{Pr}_{i,j}-{Ar}_{i,j}|}}{k}
\end{equation}
\end{itemize}
Where  $\left(i,j\right)$ are pair values in the training dataset, ${Pr}_{i,j}$ is the predicted recommendation rates, and ${Rr}_{i,j}$ is the actual recommendation rates while k is the number of recommended items. 

\begin{itemize}
\item  Recall and precision are widely used metrics to evaluate the performance of a service recommendation system. Recall is the fraction of relevant items that have been retrieved, while precision is the fraction of retrieved items that are actually relevant. The equations for recall and precision are as follows: recall = TP/(TP+FN), where TP is the number of true positives (i.e. correct classes) and FN is the number of false negatives (i.e. incorrect classes); and precision = TP/(TP+FP), where FP is the number of false positives (i.e. incorrect classes). Both metrics have an inverse relationship with each other, meaning that increasing one will decrease the other; thus, there needs to be a balance between both when training a model for optimal performance. In addition, it is important to take into account other factors such as accuracy, specificity, sensitivity, and F1 score when determining the best model for a given task.

\item  Mean Average Precision (MAP) is an evaluation metric frequently used in information retrieval tasks such as web searching and natural language processing. MAP measures the precision at which documents retrieved through an algorithm meet user expectations on average. It is calculated by taking into account both relevance and ranking order of retrieved documents; higher numbers indicate better performance while lower numbers indicate poorer performance.
\end{itemize}

\begin{equation*}
\label{Eq.19}
\begin{aligned}
AP=\sum_{k=0}^{K=n-1}{[Recalls(k)-Recalls(k+1)] 
*Precisions (k)}\\_{Recalls(n) = 0, 
Precisions(n) = 1  n= number of threshold}
\end{aligned}
\end{equation*}
Normalized Discounted Cumulative Gain (NDCG) is an additional evaluation metric to evaluate the effectiveness of recommendation systems. NDCG measures how accurately individual rankings within results lists reflect their true relevance via discounted cumulative gain scores for each result item's position within the list; higher scores indicate better alignment between actual relevance and observed rankings. NDCG takes into account not only the relevance of the recommended items but also their position in the list, giving greater importance to the most relevant items that are ranked higher. This metric is often used in industry and research to evaluate the effectiveness of different recommendation algorithms and the quality of search results.

\subsection{Performance evaluation}

The evaluation of a proposed service recommendation system approach begins by conducting a comprehensive performance analysis of various baseline methods. Subsequently, we delve into the examination of item graph structures, which are derived from multi-modal features, and their effectiveness in mitigating the cold-start problem. In this subsection, we present the experimental setup, perform the necessary experiments, and thoroughly discuss the results in accordance with the aforementioned evaluation metrics.

\subsubsection{Overall performance}

\begin{figure}
\centering
\includegraphics[width=\columnwidth]{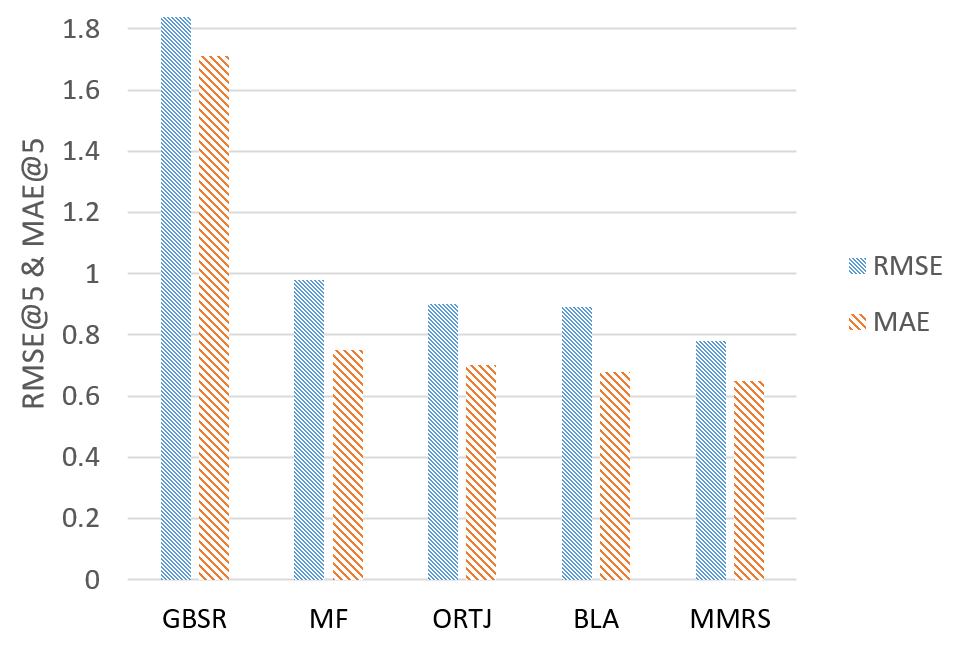}
\caption{RMSE and MAE values for Top-5 recommendation list}
\label{image_5}
\end{figure}

Upon analyzing the aforementioned results shown in Figure \ref{image_5}, GBSR method exhibits an RMSE of 1.84 and an MAE of 1.71. This implies that the predictions made by GBSR have a greater deviation from the actual values when compared to the other methods. Moving on, the Matrix Factorization (MF) method shows a significant improvement with an RMSE of 0.98 and an MAE of 0.75, indicating that this method provides more accurate predictions in comparison to GBSR. Additionally, ORTJ method exhibits better performance than both GBSR and MF, with an RMSE of 0.9 and an MAE of 0.7. These results suggest that ORTJ is a more efficient approach for making predictions with lower levels of error. Furthermore, BLA outperforms all previous methods, as it exhibits an RMSE of 0.89 and an MAE of 0.68. This implies that BLA produces more precise predictions than all other approaches. The proposed MMRS method performs the best, exhibiting an RMSE of 0.78 and an MAE of 0.65, which is the lowest RMSE and MAE among all the methods. Thus, the proposed approach is the most efficient and precise approach to make predictions in the given context.

In conclusion, the comparison of the different methods reveals that MMRS outperforms the others, followed by BLA, ORTJ, MF, and GBSR in terms of the accuracy of the predictions made. 

\begin{figure}
\centering
\includegraphics[width=\columnwidth]{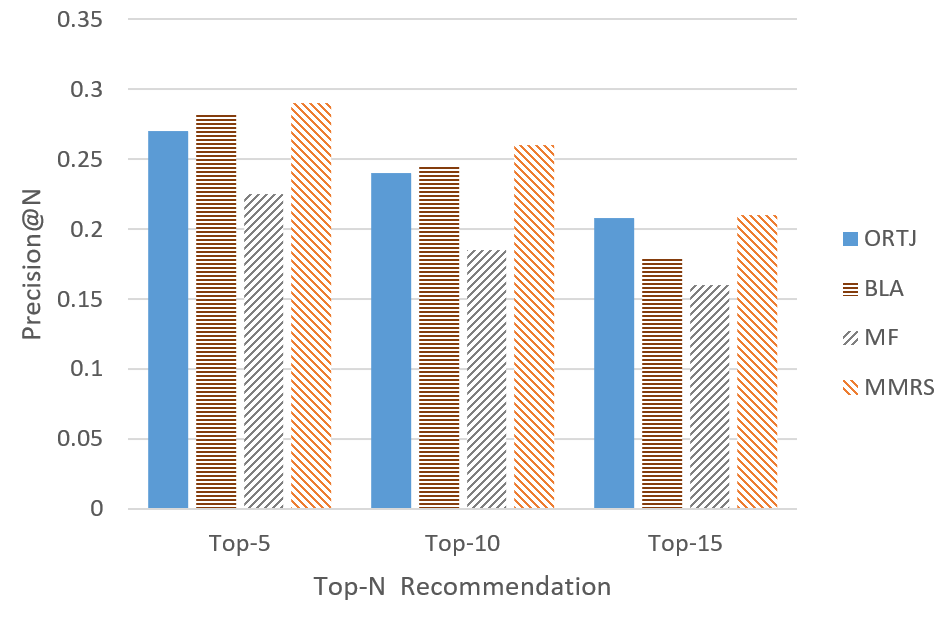}
\caption{Precision values for Top-N recommendation lis}
\label{image_6}
\end{figure}

\begin{figure}[!htb]
\centering
\includegraphics[width=\columnwidth]{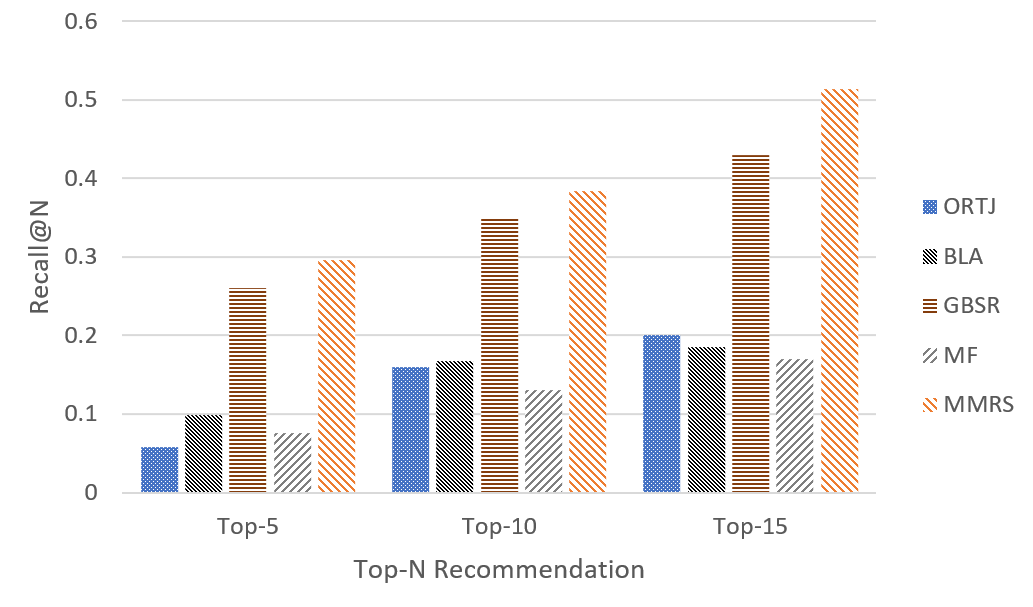}
\caption{Recall values for Top-N Recommendation}
\label{image_7}
\end{figure}

\begin{figure}[!htb]
\centering
\includegraphics[width=\columnwidth]{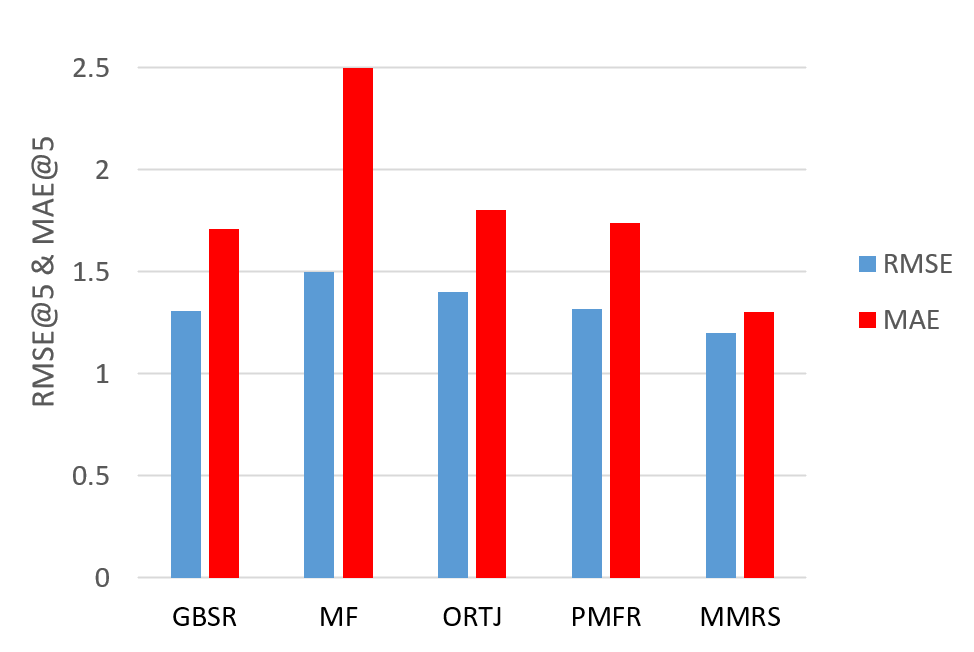}
\caption{Cold start RMSE and MAE value for top 5 recommendation}
\label{image_8}
\end{figure}

\begin{figure*}[!htbp]
\includegraphics[width=\textwidth]{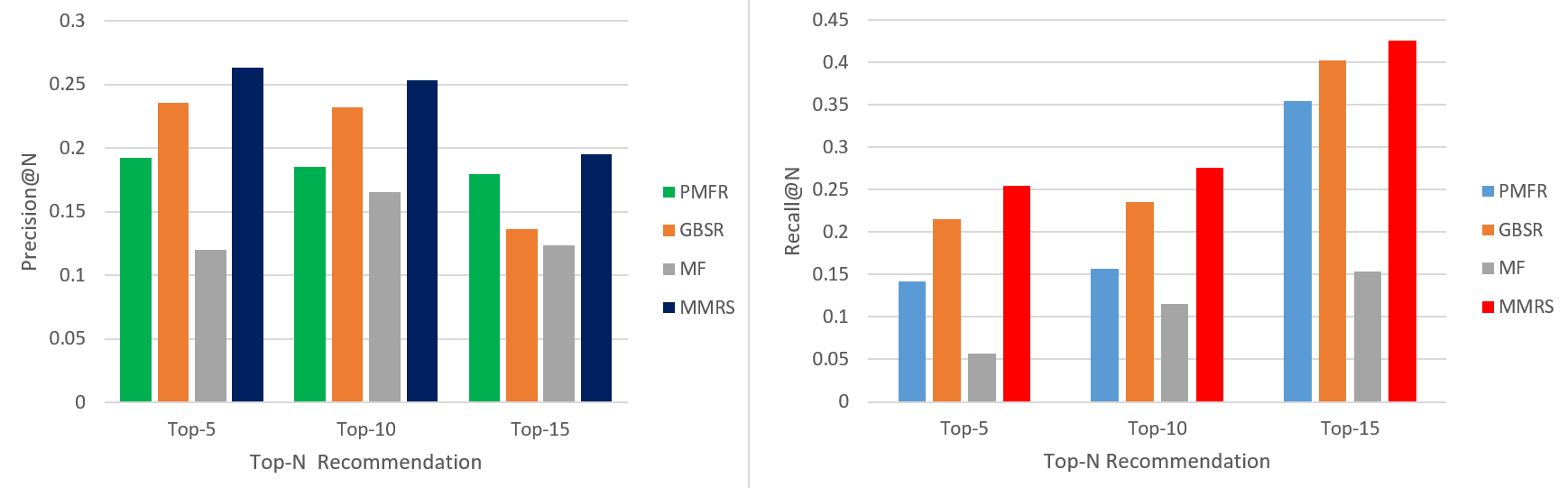}
\caption{\centering Cold-start recall and precision values} 
\label{image_9}
\end{figure*}

Figure \ref{image_6} depicts the results of the precision. Based on the obtained results, we can observe that the proposed solution MMRS consistently achieves the highest Precision values across all recommendation list sizes, indicating that it provides more accurate recommendations with a larger proportion of relevant items in the top-N recommendation lists compared to the other methods (ORTJ, BLA, and MF). For instance, at the Top-15 recommendation list, we notice that ORTJ achieves a Precision of 0.208, indicating that approximately 20.8\% of the top 15 recommended items are relevant. While BLA achieves a Precision of 0.18, suggesting that around 18\% of the top 15 recommendations are relevant. In the other hand, MF achieves the lowest Precision of 0.16, meaning that roughly 16\% of the top 15 recommended items are relevant. The proposed solution MMRS achieves a Precision of 0.21, indicating that approximately 21\% of the top 15 recommended items are relevant.

For the recall values, we can observe from the results shown in Figure \ref{image_7} that MMRS consistently achieves the highest Recall values across all recommendation list sizes, indicating that it successfully retrieves a larger proportion of relevant items within the top-N recommendation lists compared to the other methods (ORTJ, BLA, GBSR, and MF). GBSR also performs well, particularly in terms of Recall, by retrieving a substantial number of relevant items within the recommended lists.

\subsubsection{Performance in cold start setting}

The cold-start performance evaluation section is a crucial component of the proposed multi-modal service recommendation system in the realm of the SIoT. In this section, the system's ability to provide accurate and relevant recommendations to users who are new to the platform or have limited historical data is thoroughly examined. Addressing the challenges posed by the cold-start problem, which arises when there is insufficient information about users and their preferences, is essential for creating a seamless and personalized user experience. By evaluating the system's performance under such circumstances, valuable insights can be gained to enhance the recommendation algorithms and ensure optimal service recommendations for all users, regardless of their familiarity with the SIoT platform. In the cold start evaluation, we have introduced additional baseline named PMFR which is discussed in the baseline section, as it provides valuable insight on the cold start recommendation for IoT environment.

\begin{figure}[!htbp]
\centering
\includegraphics[width=\columnwidth]{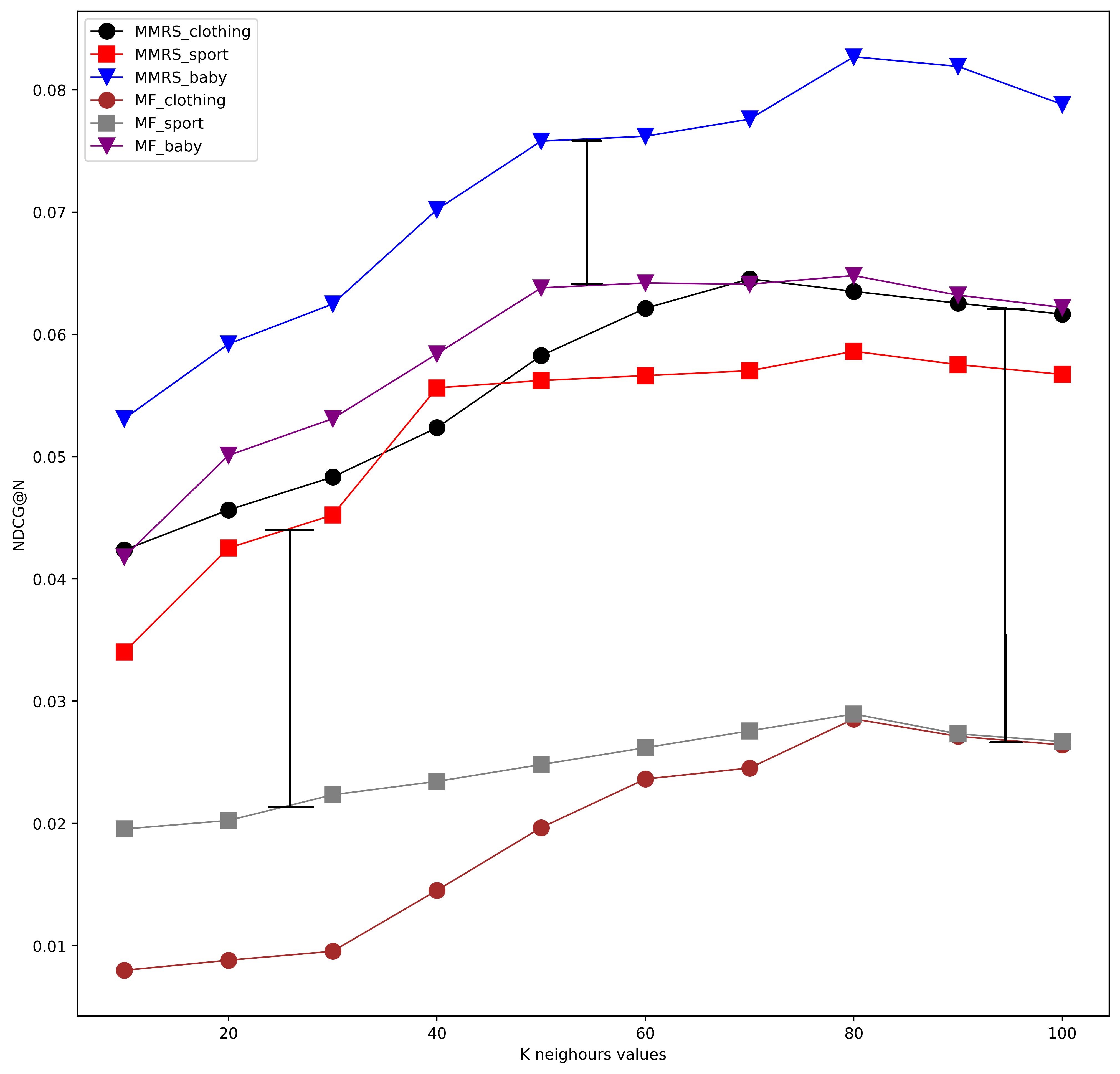}
\caption{\centering NDCG values for different datasets} 
\label{image_10}
\end{figure}

Figure \ref{image_8} presents the evaluation of RMSE and MAE performance for cold start scenarios in the top 5 recommendations. The figure demonstrates a notable decrease in performance compared to the warm start, as shown in Figure \ref{image_10}, across both metrics. However, even in the cold start situation, MMRS consistently outperforms other methods. This indicates that the proposed solution leverages multi-modal feature selection effectively, even in cases where users have limited interactions with items. The superiority of MMRS can be attributed to its reliance on item-item structure, which enables it to provide more accurate recommendations.

Likewise, the precision and recall values for cold start, depicted in Figure \ref{image_9} (a) and  Figure \ref{image_9} (b) for the top 5 to 15 recommendations respectively, demonstrate a slight decrease in performance compared to the warm start situation. Both precision and recall values are lower, indicating the significance of user-item interactions. However, MMRS continues to outperform other solutions in terms of both recall and precision, highlighting its effectiveness even in cold start scenarios.

\subsubsection{Item-item relationships factor K}

In order to gain a deeper understanding of the performance of the proposed service recommendation system, MMRS, which incorporates K-nearest neighbors (KNN), we conducted an analysis to examine the effect of the coefficient K. K represents the number of nearest neighbors considered for each item, influencing the propagation of information throughout the system. 

Figure \ref{image_10} displays the NDCG (Normalized Discounted Cumulative Gain) values for both the Matrix Factorization (MF) and our proposed method, MMRS, across different values of K ranging from 1 to 100. Our approach demonstrates a significant improvement compared to MF. Across all three datasets, particularly in the 'baby' dataset, MMRS outperforms MF in terms of NDCG. This is attributed to the sparsity of the 'baby' dataset and its abundance of related multi-modal data, allowing for more precise tailored multi-modal recommendations.

The NDCG values achieved by MMRS range from 0.02 to 0.03 for the 'sport' dataset, 0.04 to 0.06 for the 'clothing' dataset, and 0.055 to 0.1 for the 'baby' dataset. It is noteworthy that as the value of K increases, the proposed solution performs better, providing a more comprehensive understanding of the item-item relationships extracted from the multi-modal data. However, we observe a decline in performance when K exceeds 80, indicating the presence of numerous irrelevant neighbors that introduce noise and negatively impact the recommendation performance.

Overall, our analysis demonstrates the significance of carefully selecting the coefficient K in and it emphasizes the rationality of considering item relationships in multi-modal data and highlights the necessity of optimizing the number of neighbors to achieve effective recommendation performance.

\section{Conclusion}
\label{sec.Conclusion}

In this paper, we proposed a service recommendation system for Social Internet of things that uses graph structure learning to discover latent item relationships that underlie multi-modal features. The proposed system learns how different aspects of a service are related to each other, based on all the different features that can make up a service, the it develop a modality-aware graph structure learning layer that is able to effectively learn item graph structures from multi-modal features, and then combine them into a single graph. This allowed it to identify complex relationships between different kinds of features, such as how a particular service attribute might affect customer satisfaction levels. Once the graph structures are learned, graph convolutions are applied to allow each item to receive informative high-order affinities from its neighbors. Extensive experiments were conducted to validate the performance of the proposed service recommendation system using relevant and well-known performance metrics, based on real-world datasets. The results demonstrate the effectiveness of the system in providing accurate recommendations.

\bibliographystyle{IEEEtran}
\bibliography{References}

\vskip 0pt plus -1fil
\begin{IEEEbiography}[{\includegraphics[width=1in,height=1.25in,clip,keepaspectratio]{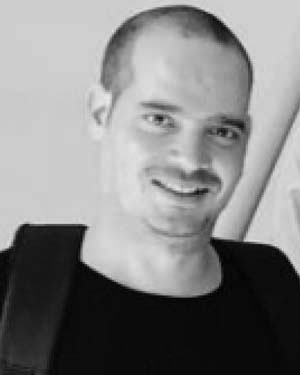}}]{Amar Khelloufi}
Received the B.S. degree (Hons.) in computer science from the Faculty of Sciences and Technology, Ziane Achour University of Djelfa, Djelfa, Algeria, in 2012, and the M.S. degree in distributed information systems from the Faculty of Sciences, University of Boumerdès, Boumerdès, Algeria, in 2014. He is currently pursuing the Ph.D. degree with the School of Computer and Communication Engineering, University of Science and Technology Beijing, Beijing, China. His current research focuses on Internet of Things, blockchain applications, edge computing, and distributed systems.
\end{IEEEbiography}

\vskip 0pt plus -1fil
\begin{IEEEbiography}[{\includegraphics[width=1in,height=1.25in,clip,keepaspectratio]{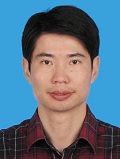}}]{Huansheng Ning}
Received his B.S. degree from Anhui University in 1996 and his Ph.D. degree from Beihang University in 2001. Now, he is a professor and vice dean of the School of Computer and Communication Engineering, University of Science and Technology Beijing, China. His current research focuses on the Internet of Things and general cyberspace. He has presided many research projects including Natural Science Foundation of China, National High Technology Research and Development Program of China (863 Project). He has published more than 200 journal/conference papers, and authored 5 books. He serves as an associate editor of IEEE Systems Journal (2013-Now), IEEE Internet of Things Journal (2014-2018), and as steering committee member of IEEE Internet of Things Journal (2016-Now).
\end{IEEEbiography}

\vskip 0pt plus -1fil
\begin{IEEEbiography}[{\includegraphics[width=1in,height=1.25in,clip,keepaspectratio]{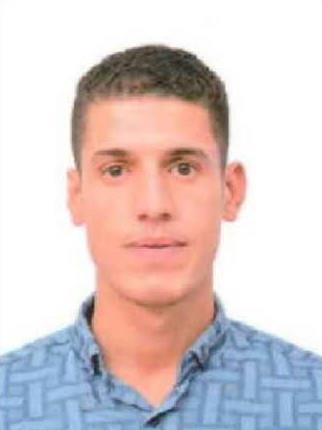}}]{Abdenacer Naouri}
He is currently a Ph.D. candidate at the University of Science and Technology Beijing China, Beijing, China. He 
received his B.S. degree in computer science from the University of Djelfa Algeria, in 2011, and the  M.Sc. degree in networking and distributed systems from the University of Laghouat Algeria, Laghouat, Algeria, in 2016. His current research interests include Cloud computing, Smart communication, machine learning, Internet of vehicles  and Internet of Things. \end{IEEEbiography}

\vskip 0pt plus -1fil
\begin{IEEEbiography}[{\includegraphics[width=1in,height=1.25in,clip,keepaspectratio]{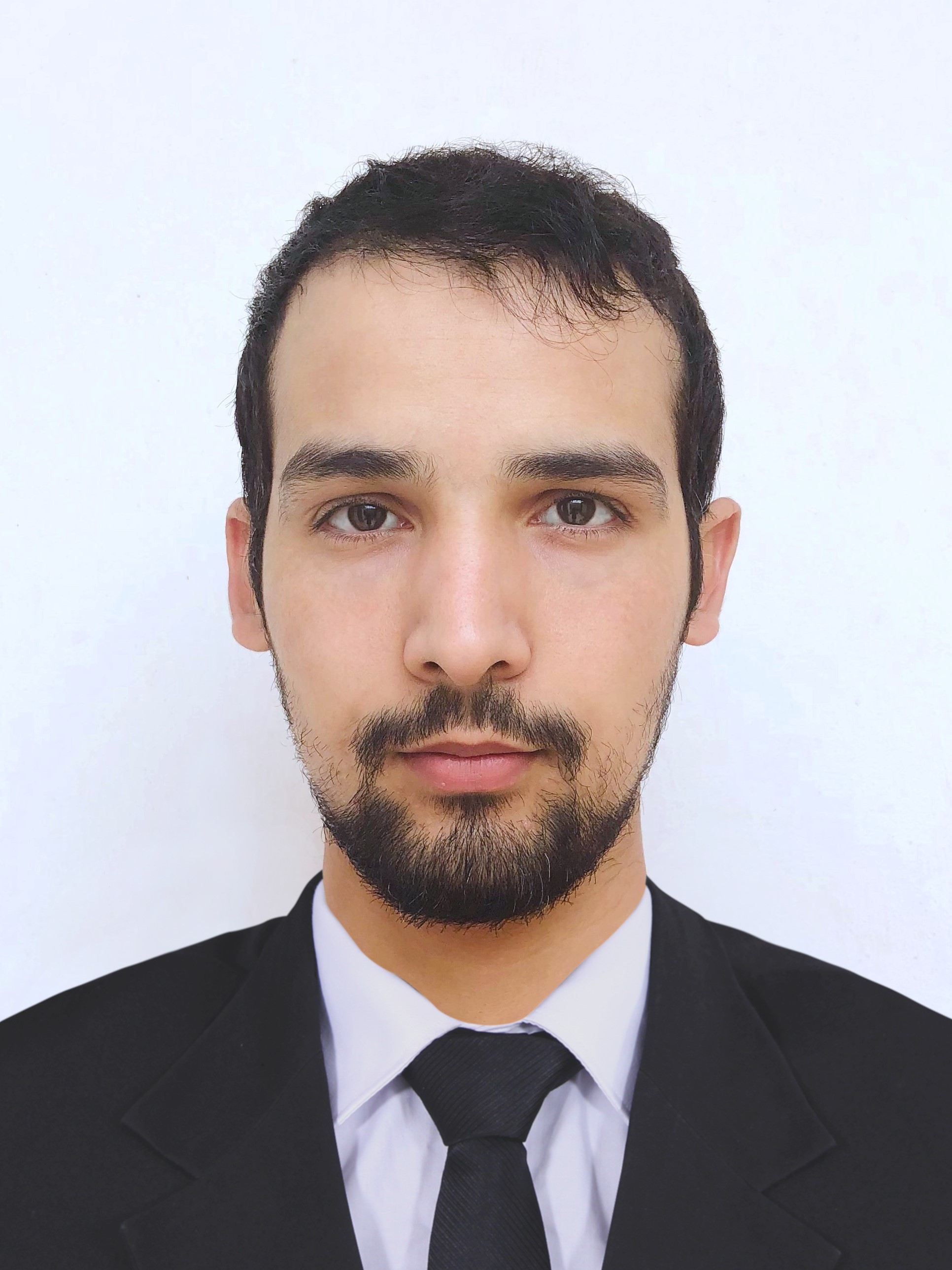}}]{Abdelkarim Ben Sada}
Received his BSc in Computer Science in 2014 from the University of Djelfa Algeria, and his MSc degree in 2016 majoring in Networking and Distributed Systems from the University of Laghouat Algeria. He is currently pursuing his PhD degree at the University of Science and Technology Beijing China. His research interests include Computer Vision, Machine Learning and Internet of Things.
\end{IEEEbiography}

\vskip 0pt plus -1fil
\begin{IEEEbiography}[{\includegraphics[width=1in,height=1.25in,clip,keepaspectratio]{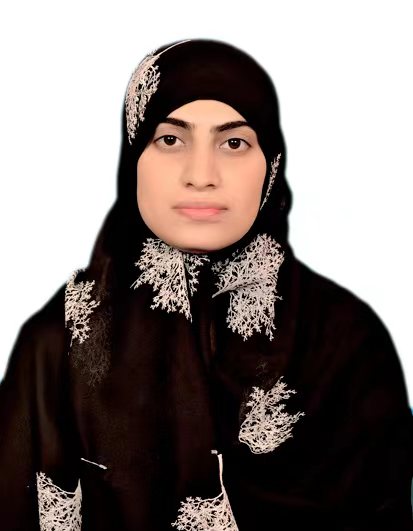}}]{Attia Qammar}
received her BS degree from Bahauddin Zakariya University and her MS from National College of Business Administration and Economics, Pakistan. Currently, she is pursuing her Ph.D. degree from the School of Computer and Communication Engineering at the University of Science and Technology Beijing, China. Her research interests include federated learning, chatbots, data security, and IoT privacy-preserving systems.
\end{IEEEbiography}

\vskip 0pt plus -1fil
\begin{IEEEbiography}[{\includegraphics[width=1in,height=1.25in,clip,keepaspectratio]{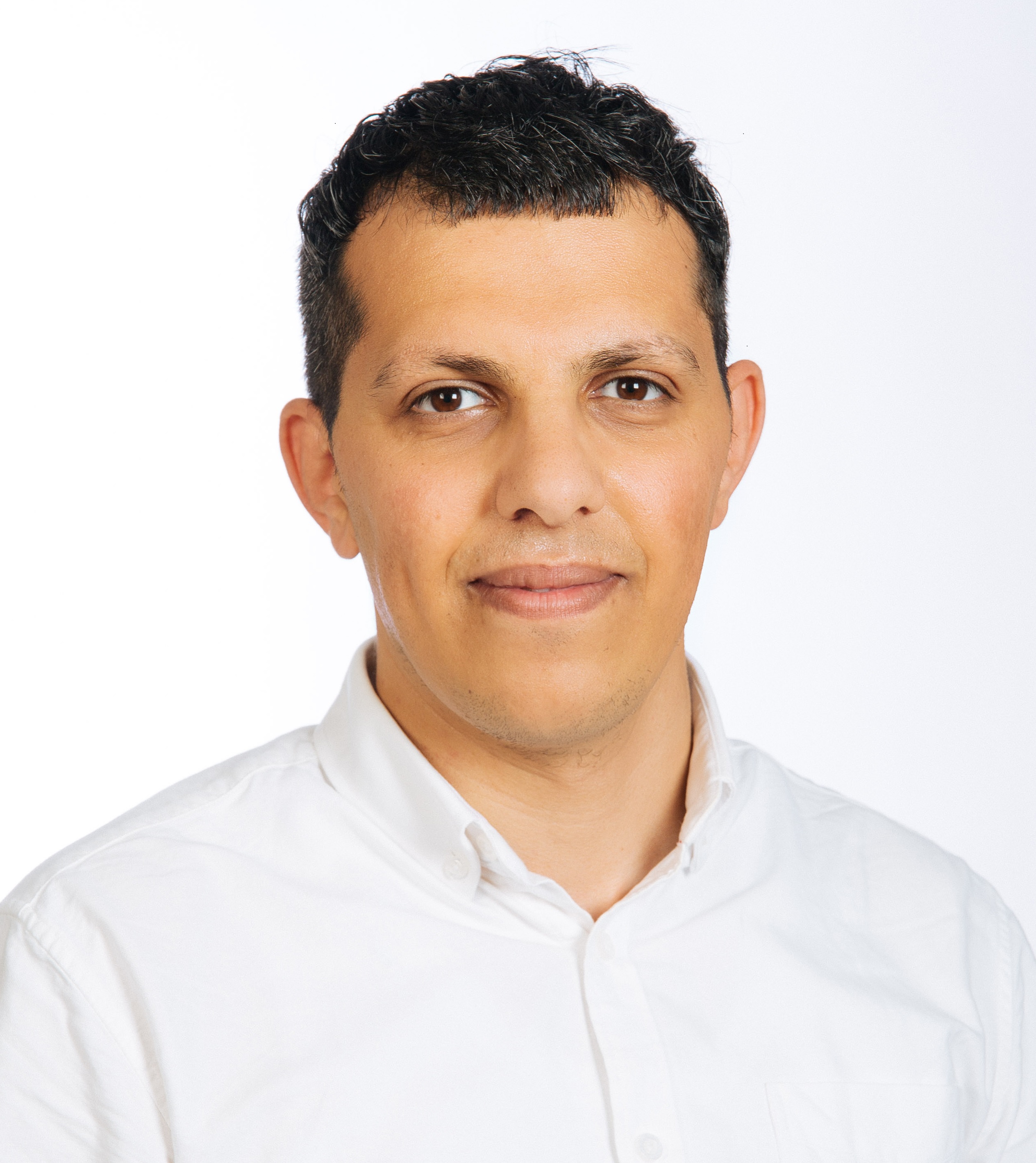}}]{Sahraoui Dhelim} is a senior postdoctoral researcher at University College Dublin, Ireland. He was a visiting researcher at Ulster University, UK (2020-2021). He obtained his PhD degree in Computer Science and Technology from the University of Science and Technology Beijing, China, in 2020. And a Master's degree in Networking and Distributed Systems from the University of Laghouat, Algeria, in 2014. And Bs degree in computer science from the University of Djelfa, in 2012. He serves as a guest editor in several reputable journals, including Electronics journal and Applied Science Journal. His research interests include Social Computing, Smart Agriculture, Deep-learning, Recommendation Systems and Intelligent Transportation Systems.
\end{IEEEbiography}

\vskip 0pt plus -1fil
\begin{IEEEbiography}[{\includegraphics[width=1in,height=1.25in,clip,keepaspectratio]{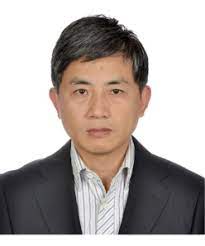}}]{Lingfeng Mao}
He is currently a professor at the School of Computer and Communication Engineering, University of Science and Technology Beijing. He received the Ph.D. degree from Peking University, Beijing, China in 2001.  He has published about 200 journal and conference papers. His research activities include reliability, modeling, and characterization of materials, devices, and systems. \end{IEEEbiography}

\newpage
\end{document}